\documentclass{article}
\usepackage{amsmath}
\usepackage{amsfonts}
\usepackage{amssymb}

\begin{document}
\bigskip
%\pagecolor{black}
%\color{white}
%
%
%
%
\begin{titlepage} \vspace{0.2in}
\begin{center} {\LARGE\bf
Revised Canonical Quantum Gravity via the Frame Fixing\\}%
\vspace*{0.8cm}
{\bf Simone Mercuri}\\ \vspace*{0.3cm}
{\bf Giovanni Montani}\\ \vspace*{1cm}
ICRA---International Center for Relativistic Astrophysics\\
Dipartimento di Fisica (G9),\\
Universit\`a  di Roma, ``La Sapienza",\\
Piazzale Aldo Moro 5, 00185 Rome, Italy.\\
e-mail: mercuri@icra.it, montani@icra.it\\
\vspace*{1.8cm}
PACS 83C  
\vspace*{1cm} \\
{\bf Abstract  \\ } \end{center} \indent
We present a new reformulation of the canonical
quantum geometrodynamics, which allows
to overcome the fundamental problem of the 
\emph{frozen formalism}
and, therefore, to construct an appropriate
Hilbert space associate to the solution of the
restated dynamics. More precisely,
to remove the ambiguity contained in the
Wheeler-DeWitt approach, with respect to the possibility
of a (3 + 1)-splitting when the space-time
is in a quantum regime, we fix the reference frame
(i.e. the lapse function and the shift vector)
by introducing the so-called kinematical action;
as a consequence the new super-Hamiltonian constraint
becomes a parabolic one and we arrive to
a Schr\"odinger-like approach for the quantum dynamics.\\
In the semiclassical limit
our theory provides General Relativity
in the presence of an additional energy-momentum density
contribution coming from no longer zero
eigenvalues of the Hamiltonian constraints;
the interpretation of these new contributions comes
out in natural way as soon as it is recognized that
the kinematical action can be recasted in
such a way it describes a pressureless, but, in general, 
non geodesic perfect fluid.
\end{titlepage}

\section{Introduction}

\label{intro}

\bigskip

The first convincing attempt to extend the methods of
quantum field theory toward the 
quantization of the gravitational field was proposed in 1967
by B. DeWitt \cite{DeW1967}. DeWitt, in his
approach, presented the implementation of the standard
canonical method to quantize the Hamiltonian constraints; 
as well known, such a procedure leads to 
limiting features which induced to search for more
appropriate reformulation of the problem. 
Until now, we do not
have a completely satisfactory theory which
unifies the fundamental principles of the
quantum mechanics with those of General Relativity, though many different
approaches exist: a review
of the key difficulties associated with the
matching of these two basic theories,
especially with respect to the concept
of time in quantum gravity, 
is presented in \cite{Ish1992} (for the more recent
loop quantum gravity and spin foam approaches, see
respectively  \cite{Thi2001} and \cite{Ori2001}).\\ 
All these different approaches have lead to develop
two main classes of theories:
one in which the time variable is determined after the
quantization procedure, and another one associated
with the choice of the ``time'' before implementing
the quantum dynamics.
We stress it exists a correlation between
these two different points of view and the 
consideration that, those 
quantization procedures, which preserve the covariance of the dynamics, 
are inconsistent with any notion of space-time slicing; indeed, if the whole metric tensor is in a full quantum
regime, then we may separate space-like objects from
time-like ones only, at most in the limiting sense of expectation
values. In this line of thinking (the incompatibility between a covariant quantization and a (3+1)-splitting), since 
the canonical method of quantization relies on
the notion of an Hamiltonian function and of its
conjugate time variables, it can be applied to gravity only
when the diffeomorphism invariance is broken.

A fully covariant approach appears well-stated in
a path integral formulation which relies on the Lagrangian function; 
this covariant point of view, initially addressed in
\cite{Har1991}, \cite{HarHaw1983}, has
found a promising development in the recent issue of the
spin foam formalism \cite{Ori2001}, in which the notion of
space-time continuity is replaced by appropriate discrete
microstructures.

The DeWitt's approach presents exactly the problem to
require the full covariance of the dynamics and preserve the
(3  + 1)-splitting of the space-time this procedure 
allows to obtain the Hamiltonian
constraints connected to the gauge invariance of the
gravitational theory
(i.e. the covariance under the space-time difffeomorphisms);
the canonical quantization of the system is then obtained by
applying the usual correspondence between the
dynamical variables and the quantum operators. When 
translating the constraints on a quantum level, 
we get from the super-Hamiltonian the so-called
Wheeler-DeWitt equation (WDE) 
\cite{MisThoWhe1973}, \cite{Kuc1973}, \cite{Kuc1981},
\cite{Ish1992} and, from the super-momentum, the invariance
of the state function under the 3-diffeomorphisms 
(J. A. Wheeler gave an important contribution to clarify
the role of a class of 3-geometries as
the fundamental variable of the theory).
 
The WDE consists of 
a functional approach in which the states of the
gravitational system are
represented by a wave functional
taken on the 3-geometries and it reflects the invariance of the
quantum dynamics under time displacements; thus, as a consequence, 
we get loss a real time dependence of the
wave functional, i.e. 
the so-called \emph{frozen formalism}, 
the most limiting outcoming features of the WDE approach
\cite{DeW1997}.

In fact, due to  the frozen formalism,
in the WDE no general procedure to turn the space of
its solutions into an Hilbert one exists; 
so any appropriate general notion, either of functional
probability either of an ``internal'' time variable 
Nevertheless, it is worth noting how,
this approach, when applied in the early cosmology problem, 
\cite{KolTur1990}, \cite{Har1988},
in which exists a good internal time
variable, the volume of the Universe,
seems to be rather expressive, especially with respect to 
the mechanism underlying the achievement of a classical Universe
after the Planckian era \cite{KirMon1997}. 

Over the last ten years the canonical quantum gravity
has found its best
improvement in the reformulation of the constraints
problem in terms of the
\emph{Ashtekar's variables}, leading to the
\emph{loop quantum gravity} theory
\cite{Ash1991}, \cite{Rov1983}, \cite{Thi2001}, which overcomes many of the
limits of the WDE, in particular it solves the
ambiguities about the construction
of an Hilbert space, and has now a wide diffusion.
However, we remark that the loop quantum gravity contains the
same ground limits of the WDE, since the (3 + 1)-splitting
of the space-time and the full covariance requirements live here till together. 
But there are reliable expectations for a link between loop
quantum gravity and the spin foam approach which,
being the latter based on a pure 4-dimensional framework,
could provide some predictivity even to the former.

Our point of view is quite different,
we think that, as stressed in
\cite{Mon2002}, many of the shortcomings of the WDE
approach are unchangeable to
a fundamental implicit \emph{ansatz,}
on which this theory is based: the possibility to speak of an Arnowitt-Deser-Misner (ADM)
formalism \cite{ArnDesMis1959}, \cite{ArnDesMis1960}, \cite{ArnDesMis1960 2},
\cite{ArnDesMis1962}, \cite{MisThoWhe1973},
when splitting the notion of space
from that one of time in a quantum regime.
Indeed, independently of the approach we are considering,
the notion of a quantum space-time  has to be associated
to the possibility of knowing metric information only in the
sense of expectation values; since to speak of a space-like
hypersurface we need to say its normal field is time-like,
then the (3 + 1) procedure make sense if translated in terms of
quantum projecting operators. But, in the canonical approaches,
the splitting is performed before quantizing and
is just in such a arbitrariness that our criticism arises.
In view of saving the quantum theory covariance, the
most rigorous approach seems to be a projecting operator
scheme which restores on a quantum level the notions of space and time.
Nevertheless the canonical approach may acquire
a precise meaning as soon as we give a physical interpretation to 
the space-time slicing even on a quantum level. To this end
we need a physical entity able to distinguish intrinsically
the nature of the time; this entity, performing a kind of
``measure'' on the quantum system, should have a certain
degree of classicality. From a more applicative point
of view, we should search for a procedure able to ensure the
existence of a time-like normal field for any choice
of the coordinates system in the quantum space-time.
We show that it is possible to joint together these two
(conceptual and applicative) scenarios, by an appropriate study
of what happens in the canonical formalism when we fix
the reference frame during the quantization scheme. 

Indeed, as shown in \cite{Mon2002}, 
fixing the slicing (in other words the reference frame)
we are able, ab initio, to distinguish between space-like and
time-like geometrical objects, so we can implement the
ADM formalism to
rewrite the action in the canonical form.
But to fix the slicing we have to
choose a particular value for the lapse function $N$
and the shift vector
$N^{i},$ and this is equivalent to 
loose the Hamiltonian constraints and, with them, the
canonical procedure of quantization.\\
The quantum field theory on a curved
background \cite{BirDav1982}, gives us a
fundamental indication about how to restore the
canonical constraints: we have to
reparameterize the action
by expressing it through the use of a generic coordinates system; 
the way toward a correct
reparameterization is provided by 
the so-called \emph{kinematical action }\cite{Kuc1981}, which
allows to achieve a satisfactory structure
for the quantum constraints.
In the case we extend this method to the
canonical quantum gravity, it is possible to get 
a ``time'' dependence of the wave
functional, with respect either to a one parameter
family of hypersurfaces,
either to a real time variable appearing on a smeared formulation 
(in this case it assumes a clear physical meaning
strictly connected with the physical
interpretation of the kinematical term, on a classical as well as
a quantum level).

The aim of our paper is to extend the
already existing (revised canonical quantum)
theory, as presented in \cite{Mon2002}, in
order to eliminate a restriction on the super-momentum
term of the kinematical action, (which is taken equal to zero
after the variation), with all its classical limit implications.\\
More precisely, 
in this paper we renfource the point of view that 
to restore the canonical
constraints after to have fixed the reference frame, it is
necessary to introduce
the kinematical action, which can be interpreted
as the action of a \emph{dust reference fluid},
which interacts with a ``gravito-electromagnetic-like'' field (GEM).
In the classical
framework the presence of the dust
is the physical consequence of fixing the
slicing, in other words it is just the
(materialized) reference fluid, in which we
describe the dynamics. The idea that, fixing the reference frame, a kind of reference fluid appears, was introduced by the works of Kucha\v{r} et al.
\cite{Kuc1991}, \cite{Kuc1992}, \cite{Kuc1992 2} and 
\cite{KucTor1991}, (an interesting discussion about the material nature of a reference frame in classical and quantum gravity associated to the problem of physical observable is provided in \cite{Rov1991}, \cite{Rov1991-1}).

Instead, on the quantum framework, 
the presence of the dust reflects
the no longer vanishing of the
eigenvalues of the quantum super-Hamiltonian
and super-momentum operators;
in particular, these eigenvalues are strictly connected to
the energy density  of
the dust and to the GEM field it fills via the influence
exerted on the world lines,
(such identifications come when taking
the classical limit of the quantum dynamics). 

To conclude, it is worth stressing that the presence of such a dust
fluid shows its effects only in those
systems which have undergone through a quantum phase
in their evolution and which
later reached a classical limit.
Thus we expect that it be possible to
observe such effects, due to the presence
of the dust, in the outcomings of the cosmological problem.

Section \ref{par1} is dedicated to the
Hamiltonian formulation
of General Relativity \cite{MisThoWhe1973},
\cite{Wal1984}, \cite{Thi2001} and, in particular, we rewrite the Einstein's equations in
terms of the canonical variables, 
without taking into account the constraints, with the aim 
of comparing them to the mean values dynamics predicted by our
revised quantum theory.\\ 
Section \ref{par2} is entirely devoted to the physical
interpretation of the kinematical action,
as viewed on a classical level, 
which is one of the building blocks of the theory,
because it allows to understand the real essence of our approach,
anticipating the outcoming of the quantum dynamics.\\ 
In Section \ref{3} we develop the quantum theory,
starting from the classical action (with the additional
kinematical term) and arriving to the functional equations;
we show that it is possible to turn the space
of the solutions of the time dependent or
``Schr\"{o}dinger-like'' equation into an Hilbert one,
with a functional
norm which allows us to define a notion of probability.
We then face the eigenvalues problem and the semiclassical
limit, which gets light on the physical meaning of
quantizing in a gauge fixing framework.\\ 
In section \ref{4} we discuss an Ehrenfest approach to the
mean value of the operator corresponding to the
classical observables; 
in the dynamical equations for the mean
values there appear additional terms in comparison
with the usual classical one, which are 
due to the particular normal ordering required by the existence 
of an Hilbert space 
and they go to zero in the classical limit.

\bigskip

\section{Hamiltonian formulation of Einstein's theory}

\label{par1}

\bigskip

To obtain the Hamiltonian constraints,
which are the starting point for the
canonical quantization of gravity, we have to write
the Einstein-Hilbert action into a (3 + 1) formulation.
To this aim, we have to perform a 
slicing of the 4-dimensional space-time, on which 
a metric tensor $g_{\mu\nu}$ is defined.

We consider a space-like hypersurface having a parametric
equation $y^{\rho}%
=y^{\rho}\left(  x^{i}\right)$
(Greek indices run from 0 to 3, while Latin ones
run from 1 to 3) and in each point
we define a 4-dimensional vector base composed by its tangent vectors 
$e_{i}^{\mu}=\partial_{i}y^{\mu}$
and by the normal unit vector $n^{\mu}$;
as just defined, 
these vectors base satisfy,
by construction, the following relations

\begin{equation}
g_{\mu\nu}e_{i}^{\mu}n^{\nu}=0,\qquad g_{\mu\nu}n^{\mu}n^{\nu}=-1.
\label{base relation}%
\end{equation}

Now if we deform this hypersurface through the whole
space-time,  via the parametric equation
$y^{\rho}=y^{\rho
}\left(  t,x^{i}\right)$, we construct a one-parameter
family of space-like hypersurfaces slicing the 4-dimensional manifold;
thus each component of the adapted base, acquiring a dependence
on the time-like parameter $t$,
becomes a vector field on the space-time.\\ 
Let us introduce the deformation vector
$N^{\mu}=\partial_{t}y^{\mu}\left(
t,x^{i}\right),$ which connects two points, 
with the same spatial coordinates,
on neighboring hypersurfaces (i.e. corresponding to values
of the parameter  $t$ and $t + dt$).\\ 
This vector field can be decomposed with respect
to the base $\left(n^{\mu},e_{i}^{\mu}\right)$, 
obtaining the following representation:

\begin{equation}
N^{\mu}=\partial_{t}y^{\rho}=Nn^{\mu}+N^{i}e_{i}^{\mu}. \label{lap shi dec}%
\end{equation}

Where $N$ and $N^i$ are, respectively, the
lapse function and the shift vector,
so this expression is known as lapse-shift
decomposition of the deformation vector.

It is easy to realize how 
the space-like hypersurfaces are characterized
by the following 3-dimensional
metric tensor  $h_{ij}=g_{\mu\nu}e_{i}^{\mu}e_{j}^{\nu}$.
Since the hypersurface
is deformed through space-time, it changes with a rate, which 
taken with respect to the label time $t$,
can be decomposed into its normal and tangential contributions 

\begin{equation}
\partial_{t}h_{ij}=-2Nk_{ij}+2\nabla_{(j}N_{i)}, \label{extrinsic curvature}%
\end{equation}

where $N_{i}=h_{ij}N^{j},$ the covariant derivative
is constructed with the
3-dimensional metric and $k_{ij}=-\nabla_{i}n_{j}$
denotes  the extrinsic curvature.

Now we define the co-base vectors
$\left(  n_{\mu},e_{\mu}^{i}\right)
,$ as follows

\begin{equation}
n_{\mu}=g_{\mu\nu}n^{\nu},\qquad e_{\mu}^{i}=h^{ij}g_{\mu\nu}e_{j}^{\nu},%
\end{equation}

where $h^{ij}$ is the
inverse 3-metric: $h^{ij}h_{jk}=\delta_{k}^{i}.$ By the
second of the above relation
we obtain $e_{\mu}^{i}e_{j}^{\mu}=\delta_{j}^{i}.$

The explicit expression for the 4-metric tensor
$g_{\mu \nu}$ and its
inverse $g^{\mu\nu}$ assume, in the system $\left(  t,x^{i}\right)$, respectively, the form:

\begin{equation}
g_{\mu\nu}=\left(
\begin{array}
[c]{cc}%
N_{i}N^{i}-N^{2} & N_{i}\\
N_{i} & h_{ij}%
\end{array}
\right),\qquad g^{\mu\nu}=\left(
\begin{array}
[c]{cc}%
-\dfrac{1}{N^{2}} & \dfrac{N^{i}}{N^{2}}\\
\dfrac{N^{i}}{N^{2}} & h^{ij}-\dfrac{N^{i}N^{j}}{N^{2}}%
\end{array}
\right). \label{metric}%
\end{equation}

Moreover, the normal vector $n^{\mu}$
has the following components $\left(
\dfrac{1}{N},-\dfrac{N^{i}}{N}\right)$, and
this implies that the covariant
normal vector be $n_{\mu}=\left(  -N,0\right)$; below 
we will use to indicate the components of the vectors
in the system $\left(  t,x^{i}\right)  ,$ with Greek
barred indices as: $\bar{\mu},$ $\bar{\nu}$,
$\bar{\rho}.....$ We
also note that in this system of coordinates, 
the square root of the determinant of the metric
tensor assumes the form $\sqrt{-g}=N\sqrt{h}.$

It is possible to show that the Einstein-Hilbert action
can be rewritten as follows \cite{ArnDesMis1962}, \cite{Wal1984}, \cite{Thi2001}:

\begin{equation}
S=\int\limits_{\Sigma^{3}\times\Re}dtd^{3}xN\sqrt{h}\left(  ^{\left(
3\right)  }R+k_{ij}k^{ij}-k^{2}\right), \label{azione grav}%
\end{equation}

which is the most appropriate to construct the ``ADM action''
for the gravitational field.\\ 
Now, defining the conjugate momenta to the
dynamical variables, which are the
component of the 3-metric tensor, we can rewrite the
gravitational action in its Hamiltonian form.
The gravitational Lagrangian $L^{g}$ does not contain the
time derivative of the lapse function $N$ and of the shift
vector $N^{i},$ so
their conjugate momenta are identically zero and the
Lagrangian is said singular.
Summarizing, we have for the conjugate momenta:

\begin{align}
&  p^{ij}\left(  t,x^{i}\right)  =\dfrac{\partial L^{g}}{\partial\left(
\partial_{t}h_{ij}\right)  }=\sqrt{h}\left(  k^{ij}-kh^{ij}\right)  ,\\
\pi\left(  t,x^{i}\right)   &  =\dfrac{\partial L^{g}}{\partial\left(
\partial_{t}N\right)  }=0,\qquad\pi_{i}\left(  t,x^{i}\right)  =\dfrac
{\partial L^{g}}{\partial\left(  \partial_{t}N^{i}\right)  }=0.
\end{align}
\qquad

By the above definition,
we can perform the Legendre dual transformation and,
with few algebra, then obtaining the below final form for
the gravitational action \cite{Thi2001}:

\begin{align}
S^{g}\left(  h_{ij},p^{kl},N,N^{a},\pi,\pi_{b}\right)   &  =\int
\limits_{\Sigma^{3}\times\Re}dtd^{3}x\left\{  p^{ij}\partial_{t}h_{ij}%
+\pi\partial_{t}N+\pi_{k}\partial_{t}N^{k}\right. \label{az grav}\\
&  -\left.  \left(  \lambda\pi+\lambda^{j}\pi_{j}+NH^{g}+N^{i}H_{i}%
^{g}\right)  \right\}
\end{align}

where the so called super-Hamiltonian
$H^{g}$ and super-momentum $H_{i}^{g},$ read respectively as

\begin{equation}
H^{g}=G_{ijkl}p^{ij}p^{kl}-\sqrt{h}^{\left(  3\right)  }R,\qquad H_{i}%
^{g}=-2\nabla_{j}p_{i}^{j},
\end{equation}

where 
(using geometrical units)
$G_{ijkl}=\dfrac{1}{2\sqrt{h}}\left(
h_{ik}h_{jl}+h_{il}h_{jk}-h_{ij}h_{kl}\right)  $
is the so-called super-metric.

Now, before calculating the other dynamical equations, we
want to add to this picture, also
a matter field, which, for simplicity, is represented by
a self-interacting
scalar field $\phi.$ This lead us to the following
expression for the action
of the gravitational and matter field:

\begin{align}
S^{g\phi}  &  =\int\limits_{\Sigma^{3}\times\Re}dtd^{3}x\left\{
p^{ij}\partial_{t}h_{ij}+\pi\partial_{t}N+\pi_{k}\partial_{t}N^{k}+p_{\phi
}\partial_{t}\phi\right. \nonumber\\
&  -\left.  \left(  \lambda\pi+\lambda^{j}\pi_{j}+N\left(  H^{g}+H^{\phi
}\right)  +N^{i}\left(  H_{i}^{g}+H_{i}^{\phi}\right)  \right)  \right\}
\label{grav-scal action}%
\end{align}

where the Hamiltonian terms $H^{\phi}$ and $H_{i}^{\phi}$
read explicitly as:

\begin{equation}
H^{\phi}=\dfrac{1}{2\sqrt{h}}p_{\phi}^{2}+\dfrac{\sqrt{h}}{2}h^{ij}%
\partial_{i}\phi\partial_{j}\phi+\sqrt{h}V\left(  \phi\right)  \qquad
H_{i}^{\phi}=p_{\phi}\partial_{i}\phi
\end{equation}

and $V\left(\phi\right)$ denotes the self-interaction
potential energy.

Varying the action (\ref{grav-scal action}) with respect to the
Lagrange multipliers $\lambda$ and $\lambda_{i}$, we obtain
the first class constraints:

\begin{equation}
\pi=0,\qquad \pi_{k}=0 \label{ham con};
\end{equation}

to assure that the dynamics is consistent, i.e. the Poisson parenteses between the constraints
and the Hamiltonian be zero, we have to require that the second class constraints

\begin{equation}
H^{g}+H^{\phi}=0,\qquad H_{i}^{g}+H_{i}^{\phi}=0, \label{ham con2}
\end{equation}

be satisfied.

Moreover varying the action with respect the two
conjugate momenta $\pi$ and $\pi_{i}$, we obtain the two
equations

\begin{equation}
\partial_{t}N=\lambda,\qquad \partial_{t}N^{i}=\lambda^{i}, \label{eq lap shi}
\end{equation}

which ensure that the trajectories of the lapse
function and of the shift
vector in the phase space are completely arbitrarily.

The action (\ref{grav-scal action}) has to be varied with respect to
all the dynamical variables and this gives
us the Hamiltonian equations for the scalar and the gravitational field, which take the form:

\begin{equation}
\dfrac{d}{dt}h_{ab}=2NG_{abkl}p^{kl}+2\nabla_{(a}N_{b)}, \label{eq per h}%
\end{equation}%

\begin{align}
\dfrac{d}{dt}p^{ab}  &  =\dfrac{1}{2}N\dfrac{h^{ab}}{\sqrt{h}}\left(
p^{ij}p_{ij}-\dfrac{1}{2}p^{2}\right)  -\dfrac{2N}{\sqrt{h}}\left(
p^{ai}p_{i}^{b}-\dfrac{1}{2}pp^{ab}\right)  +\nonumber\\
&  -N\sqrt{h}\left(  ^{\left(  3\right)  }R^{ab}-\dfrac{1}{2}^{\left(
3\right)  }Rh^{ab}\right)  +\nonumber\\
&  +\sqrt{h}\left(  \nabla^{a}\nabla^{b}N-h^{ab}\nabla^{i}\nabla_{i}N\right)
+\nonumber\\
&  -2\nabla_{i}\left(  p^{i(a}N^{b)}\right)  +\nabla_{i}\left(  N^{i}%
p^{ab}\right)  +\nonumber\\
&  +\dfrac{N}{4\sqrt{h}}h^{ab}p_{\phi}^{2}-\dfrac{N}{2}\sqrt{h}h^{ab}\left(
\dfrac{1}{2}h^{ij}\partial_{i}\phi\partial_{j}\phi+V\left(  \phi\right)
\right),  \label{eq per p}%
\end{align}%

\begin{equation}
\dfrac{d}{dt}\phi=\dfrac{N}{\sqrt{h}}p_{\phi}+N^{i}\partial_{i}\phi,
\end{equation}%

\begin{align}
\dfrac{d}{dt}p_{\phi}  &  =N\sqrt{h}h^{ij}\partial_{i}\partial_{j}%
\phi+\partial_{j}\left(  N\sqrt{h}h^{ij}\right)  \partial_{i}\phi+\nonumber\\
&  -N\sqrt{h}\dfrac{\partial V\left(  \phi\right)  }{\partial\phi}%
+\partial_{i}\left(  N^{i}p_{\phi}\right).
\end{align}

The complete dynamics of the coupled gravito-scalar system is represented by the above dynamical equations together with equation (\ref{eq lap
shi}) and the first and second class constraints (\ref{ham con}) and (\ref{ham con2}),
which tell us we can not choose
the fields and their conjugate momenta arbitrarily.

$\bigskip$

\section{Physical interpretation of the ``kinematical action''}

\label{par2}

\bigskip

We have introduced in the previous
section the lapse-shift decomposition of
the deformation vector (\ref{lap shi dec}). It is worth noting that we can
obtain such equation varying an action built to this aim.
It is the so-called kinematical action
and takes the following form:

\begin{equation}
S=\int\limits_{\Sigma^{3}\times\Re}dtd^{3}x\left(  p_{\mu}\partial_{t}y^{\mu
}-Np_{\mu}n^{\mu}-N^{i}p_{\mu}e_{i}^{\mu}\right).  \label{kin act}%
\end{equation}

If we now vary the action (\ref{kin act})
with respect to the dynamical
variables $p_{\mu}$ and $y^{\mu}$, and we
put these two variations equal to
zero, we obtain respectively:

\begin{equation}
\partial_{t}y^{\mu}=Nn^{\mu}+N^{i}\partial_{i}y^{\mu},
\qquad \partial_{t}p_{\mu}=-Np_{\rho}\partial_{\mu}n^{\rho
}+\partial_{i}\left(  N^{i}p_{\mu}\right). \label{eq din}%
\end{equation}

The first one of such equations is the lapse-shift
decomposition of the deformation
vector, while the second one provides the dynamical
evolution for $p_{\mu},$ which is
the conjugate momenta to the vector $y^{\rho}.$

The kinematical action is used in quantum
field theory on curved space-time, in order to
reparameterize the field action \cite{BirDav1982},
\cite{Kuc1981}, but it will
be clear in the next section how, in our approach,
it plays an important role
also in the
reformulation of the canonical quantum gravity.

In this section we want to investigate the
physical meaning of the
``kinematical term'', which will outline 
either the main aspects of our
reformulation of the canonical quantum gravity,
either the meaning of the
reparameterization in quantum field on curved space.

To get the searched physical insight, 
let us rewrite the equations (\ref{eq din})
in a covariant form. To this aim
we recall to denote the coordinates $(t,x^{i})$ by barred Greek
indices: $\overline{\mu
},\overline{\nu},\overline{\rho}....$ and we also
remark that the following
relations take place: $\partial_{t}=\partial_{t}y^{\mu}\partial_{\mu},$
$\partial_{i}=\partial_{i}y^{\mu}\partial_{\mu},$\ $n^{\overline{\mu}}%
\partial_{\overline{\mu}}=n^{\mu}\partial_{\mu}.$\\
Now remembering that the
normal vector $n^{\mu}$ has components 
$n^{\overline{\mu}}\equiv\left(
\dfrac{1}{N},-\dfrac{N^{i}}{N}\right)  $
in the system $\left(  t,x^{i}%
\right)  ,$ it is possible to rewrite the first one
of equations (\ref{eq
din}) in the following form:
$n^{\mu}=n^{\overline{\rho}}\partial
_{\overline{\rho}}y^{\mu};$ this equation ensures that, after the variation
$n^{\mu}$ is a real unit time-like vector, i.e.

\begin{equation}
g_{\mu\nu}n^{\mu}n^{\nu}=g_{\mu\nu}n^{\overline{\rho}}\partial_{\overline
{\rho}}y^{\mu}n^{\overline{\sigma}}\partial_{\overline{\sigma}}y^{\nu
}=g_{\overline{\rho}\overline{\sigma}}n^{\overline{\rho}}n^{\overline{\sigma}%
}=-1, \label{eq nor vec}%
\end{equation}

the last equality being true by construction
of $g_{\overline{\mu}%
\overline{\nu}}$ and $n^{\overline{\mu}}.$
Moreover, since $n^{\mu }$ is in any system of
coordinates normal to the hypersurfaces $\Sigma^{3},$ then
we see how the use of the kinematical action allows to
overcome the ambiguity in the existence of a real
time-like normal vector field, we have spoken about in the introduction
of this paper. 

Now using the relations
$\partial_{t}=\partial_{t}y^{\mu}\partial_{\mu},$
$\partial_{i}=\partial_{i}y^{\mu}\partial_{\mu},
$\ $n^{\overline{\mu}}%
\partial_{\overline{\mu}}=n^{\mu}\partial_{\mu}$
and the first one of equations (\ref{eq din}),
we may rewrite the second kinematical equation,
concerning the momentum dynamics  as follows:

\begin{equation}
n^{\rho}\left[  \partial_{\rho}\left(  Np_{\mu}\right)  -\partial_{\mu}\left(
Np_{\rho}\right)  \right]  =-\partial_{\mu}\left(  Np_{\nu}n^{\nu}\right)
+p_{\mu}\left(  n^{\rho}\partial_{\rho}N+\partial_{i}N^{i}\right)  ;
\label{eq4}%
\end{equation}

we note that $p_{\mu}$ is not a vector, but it is
a vector density of weight 1/2; thus we can rewrite it
as $p_{\mu}=-\sqrt{h}\varepsilon\pi_{\mu}$,
where $\varepsilon$ is a real 3-scalar and
$\pi_{\mu}$ is a vector, such that 
it satisfies the relation
$n^{\mu}\pi_{\mu}=-1.$ Using this new
expression for $p_{\mu}$, equation (\ref{eq4}) rewrites:

\begin{equation}
\varepsilon n^{\rho}\left(  \partial_{\rho}\pi_{\mu}-\partial_{\mu}\pi_{\rho
}\right)  =-\pi_{\mu}\frac{1}{\sqrt{-\overline{g}}}\partial_{\rho}\left(
\sqrt{-\overline{g}}\varepsilon n^{\rho}\right)  , \label{eq5}%
\end{equation}

which covariantly reads

\begin{equation}
\varepsilon n^{\rho}\left(  \nabla_{\rho}\pi_{\mu}-\nabla_{\mu}\pi_{\rho
}\right)  +\pi_{\mu}\nabla_{\rho}\left(  \varepsilon n^{\rho}\right)  =0.
\label{eq6}%
\end{equation}

Then, multiplying equation (\ref{eq6}) for $n^{\mu}$, 
we get

\begin{equation}
\nabla_{\rho}\left(  \varepsilon n^{\rho}\right)  =0. \label{eq con}%
\end{equation}

A perfect fluid, having entropy density $\sigma$
and 4-velocity $u_{\mu},$
satisfies the equation
$\nabla_{\mu}\left(  \sigma u^{\mu}\right)  =0$, but
for a dust case the density of entropy is proportional
to the density of energy
$\left(  \sigma\propto\varepsilon\right)  $,
so that equation (\ref{eq con}) is the
one for a dust fluid of density of energy $\varepsilon$
and 4-velocity $n_{\mu}.$

Now, using equation (\ref{eq con}), we can rewrite the
relation (\ref{eq6}) as%

\begin{equation}
n^{\rho}\left(  \nabla_{\rho}\pi_{\mu}-\nabla_{\mu}\pi_{\rho}\right)  =0.
\label{eq7}%
\end{equation}

Setting now
$\pi_{\mu}=n_{\mu}+s_{\mu}$, with $n^{\mu}s_{\mu}=0$, from 
above, we arrive to

\begin{equation}
n^{\rho}\nabla_{\rho}n_{\mu}=n^{\rho}\left(  \nabla_{\mu}s_{\rho}-\nabla
_{\rho}s_{\mu}\right)  =\gamma n^{\rho}F_{\mu\rho}, \label{eq8}%
\end{equation}

with $s_{\rho}=\gamma A_{\rho}$,
where $\gamma$ is a constant and
$F_{\mu\rho
}=\nabla_{\mu}A_{\rho}-\nabla_{\rho}A_{\mu}$
(obviously $n^{\rho}A_{\rho}=0$).

Thus equation (\ref{eq8}), together with (\ref{eq con})
are the field equations of a dust fluid with density of
energy $\varepsilon$, whose 4-velocity $n^{\mu}$ is tangent 
to a space-time curve associated to the presence of an
''electromagnetic-like'' field
(say a \emph{gravito-electromagnetic field}). So, on a classical level, the kinematical action
is equivalent to the action of such a dust fluid and,
in this sense, it is upgrated
from its geometrical nature to a physical state.

The condition $n^{\rho}A_{\rho}=0$ can be written
in the system $\left(
t,x^{i}\right)  $ as
$n_{\overline{\rho}}A^{\overline{\rho}}=0$,
from which it follows 
$A^{\overline{0}}=0$ and this means that
in the fluid reference we have to do with a gauge condition 
such that $A^{\mu}\equiv\left(  0,\underline{A}\right)  ,$ i.e. 
with a simple 3-vector potential for the gravito-electromagnetic field.

Now let us come back to the kinematical action (\ref{kin act}):
varying it with respect to $N$ and $N^{i}$,
we obtain the corresponding super-Hamiltonian and
super-momentum of the kinematical term:

\begin{equation}
H^{k}=p_{\mu}n^{\mu},\qquad H_{i}^{k}=p_{\mu}e_{i}^{\mu}, \label{sup kin}%
\end{equation}

Using the definitions above introduced for
$p_{\mu}$ and $s_{\mu}$ we have:

\begin{equation}
H^{k}=\sqrt{h}\varepsilon,\qquad H_{i}^{k}=-\sqrt{h}\varepsilon\gamma A_{\mu
}e_{i}^{\mu}. \label{ham mom}%
\end{equation}

It is clear that
$A_{\mu}e_{i}^{\mu}=A_{\mu}\dfrac{\partial y^{\mu}}{\partial
x^{i}}$ is a transformation of coordinates from
the generic system $y^{\mu}$
to the system of the hypersurface, that is the
one which we have before indicated with
barred indices. So we write $A_{\mu}e_{i}^{\mu}=A_{i}$, 
that is we introduce the projection of
the field $A_{\mu}$ on the spatial hypersurfaces.

So equations
(\ref{ham mom}) rewrites as:

\begin{equation}
H^{k}=\sqrt{h}\varepsilon,\qquad H_{i}^{k}=-\sqrt{h}\varepsilon\gamma A_{i}.
\label{ham mom2}%
\end{equation}

In \cite{Mon2002} is shown that the energy-momentum
tensor of the dust is
orthogonal to the hypersurfaces $\Sigma^{3};$
this is the reason why it contributes only to the
super-Hamiltonian, by its energy density.
Moreover, it is possible to show,
via a simple model, why the presence of the field
$A_{\mu}$ has, instead, effects only on the super-momentum.
To this end, let us consider an interaction
between a current $j^{\mu}$ and a field $B_{\mu}$;
then the Hamiltonian of interaction will be:

\begin{equation}
H_{int}=\int d^{4}x\sqrt{-g}j^{\mu}B_{\mu}. \label{ham int}%
\end{equation}

Since $H_{int}$ is obviously a scalar, we can
rewrite it in the system of
coordinates with barred indices, as follows 

\begin{equation}
H_{int}=\int d^{4}\overline{x}N\sqrt{h}j^{\overline{\mu}}B_{\overline{\mu}};
\label{ham int bar}%
\end{equation}

taking now $j^{\overline{\mu}}=\varepsilon n^{\overline{\mu}}$
(current of matter) and
$B_{\overline{\mu}}=\gamma A_{\overline{\mu}},$ we have,
remembering also
that $n^{\overline{\mu}}\equiv\left(  \dfrac{1}{N}%
,-\dfrac{N^{i}}{N}\right)  ,$

\begin{equation}
H_{int}=\int d^{4}\overline{x}\sqrt{h}\varepsilon\gamma\left(  A_{\overline
{0}}-N^{\overline{i}}A_{\overline{i}}\right)  . \label{ham int fin}%
\end{equation}

This expression no more depends on the lapse function,
so that it does not
contribute to the super-Hamiltonian, while
the contribution to the super-momentum
is just the one in equation (\ref{ham mom}).

Above we have introduced the projection of the field $A_{\mu}$
on the spatial hypersurfaces, i.e.
$A_{i}=A_{\mu}e_{i}^{\mu};$ this is of course a
simple transformation of coordinates, but
it does not assure $A_{i}$ is a 3-vector.
To show this, we define $A^{i}=A^{\mu}e_{\mu}^{i};$
it is worth noting that it is 
not a transformation of coordinates, but
this choice on how to project the
contravariant 4-vector $A^{\mu}$, is sufficient to show
that $A_{i}=h_{ik}%
A^{k},$ which ensures $A_{i}$ is a 3-vector
on the hypersurfaces, which lowers
and raises its index by the induced 3-metric.\\ 
In fact starting from the expression of $A_{i}$
and recalling that $e_{i}^{\mu }=h_{ik}g^{\mu\nu}e_{\nu}^{k},$,
we can write:

\begin{equation}
A_{i}=A_{\mu}e_{i}^{\mu}=A_{\mu}h_{ik}g^{\mu\nu}e_{\nu}^{k}=h_{ik}A^{\nu
}e_{\nu}^{i}=h_{ik}A^{k}, \label{3-vector}%
\end{equation}

where, in the last equality, we have used the
definition of $A^{k}.$

To conclude this section, we want to study
the behaviors of $\varepsilon$ and
$A_{i};$ to this end we start from equations
(\ref{eq din}), multiplying the
second one by $n^{\mu},$ and remembering that
$n^{\mu}\partial_{\mu}%
=n^{\overline{\mu}}\partial_{\overline{\mu}},$ we arrive to 

\begin{equation}
\partial_{t}\left(  \sqrt{h}\varepsilon\right)  -\partial_{i}\left(  \sqrt
{h}\varepsilon N^{i}\right)  =0; \label{eq e}%
\end{equation}

Moreover, by multiplying the second one
with $e_{i}^{\mu}$ and considering
also the first kinematical equation, we get  
an expression of the form: 

\begin{equation}
\partial_{t}\left(  \sqrt{h}\varepsilon\gamma A_{i}\right)  -\partial
_{k}\left(  \sqrt{h}\varepsilon\gamma N^{k}A_{i}\right)  =\sqrt{h}%
\varepsilon\gamma A_{k}\partial_{i}N^{k}-\sqrt{h}\varepsilon\partial_{i}N.
\label{eq A}%
\end{equation}

To treat these two equations (\ref{eq e}) and (\ref{eq A})
in a general reference frame, it is a very difficult task,
 but it becomes very simple in a
synchronous reference, where $N=1$ and $N^{i}=0;$
in this particular case we have:

\begin{equation}
\partial_{t}\left(  \sqrt{h}\varepsilon\right)  =0,\qquad\partial_{t}\left(
\sqrt{h}\varepsilon\gamma A_{i}\right)  =0.
\end{equation}

The first one of the above equations means that
$\sqrt{h}\varepsilon
=-\omega\left(  x^{i}\right)  $ where
$\omega$ is a scalar density of weight
1/2, which depends only on $x^{i};$ we note that
$\varepsilon=-\dfrac
{\omega\left(  x^{i}\right)  }{\sqrt{h}}$, this means
$\varepsilon$ is the
density of energy of a non relativistic dust.
While from the second one we obtain 
$\gamma A_{i}\omega\left(  x^{i}\right)
=-k_{i}\left(  x^{k}\right)  ,$ which
is a 3-vector density of weight 1/2 and
depends only on $x^{i}$
(we have to do with a simple magnetic term).
It is
clear that we can now write the super-Hamiltonian
and super-momentum of the kinematical term as follows 

\begin{equation}
H^{k}=-\omega\left(  x^{l}\right)\qquad H_{i}^{k}=-k_{i}\left(
x^{l}\right)  \label{raffaella}
\end{equation}

We will return on the above expression
in the next section, when treating the
eigenvalues problem and the classical limit of
the quantized theory; indeed 
we will find a
connection between the density of energy of the dust and
the eigenvalue of the super-Hamiltonian operator
as well as between the eigenvalues of the
super-momentum operator and the presence of the field $A_{i}.$

$\bigskip$

\section{Canonical quantization of the model}

\label{3}

$\bigskip$

Our reformulation of the canonical
quantum gravity is based on a fundamental
criticism about the possibility to
speak of a unit time-like normal field and
of space-like hypersurfaces, which are
at the ground of the ADM formalism,
when referring to a quantum space-time;
in fact, in this case, either the
time-like nature of a vector field, either the space-like
nature of the hypersurfaces can be recognized at most in
average sense, i.e. with respect to
expectation values. This consideration makes
extremely ambiguous to apply the
3+1 splitting on a quantum level and leads us to claim
that the canonical quantization of gravity has sense
only when referred to a fixed slicing, or in
other words, when referred to a fixed reference frame,
i.e. only after the notion of space and time are
physically distinguishable. To fix the slicing we
have to choose a particular family of hypersurfaces
and this means we have to fix
the lapse function $N$ and the shift vector $N^{i}.$
However, so doing, we
loose the Hamiltonian constraints (\ref{ham con}), (\ref{ham
con2}) and, with
them, a standard procedure to quantize the
dynamics of the system;
as a solution to this problem, we propose to
introduce, like in the fixed background field theory,
the kinematical action \cite{Kuc1981} to the total gravity-matter one,
to reparameterize the action;
hence we obtain:
\begin{align}
S^{g\phi k} &  =\int\limits_{\Sigma^{3}\times\Re}dtd^{3}x\left\{
p^{ij}\partial_{t}h_{ij}+\pi\partial_{t}N+\pi_{k}\partial_{t}N^{k}+p_{\phi
}\partial_{t}\phi+p_{\mu}\partial_{t}y^{\mu}\right.  +\nonumber\\
&  -\left.  \left(  \lambda\pi+\lambda^{i}\pi_{i}+N\left(  H^{g}+H^{\phi
}+H^{k}\right)  +N^{i}\left(  H_{i}^{g}+H_{i}^{\phi}+H_{i}^{k}\right)
\right)  \right\}  .\label{azione totale}%
\end{align}

Now the lapse function $N$ and the shift
vector $N^{i}$ are to be regarded as
dynamical variables, obtaining in this way
the new Hamiltonian constraints, i.e.
\begin{align}
\pi &  =0,\qquad\pi_{k}=0,\label{new ham con}\\
H^{g}+H^{\phi}+H^{k}  &  =0,\qquad H_{i}^{g}+H_{i}^{\phi}+H_{i}^{k}=0.
\label{new ham con2}%
\end{align}

We have, also, introduced the new dynamical field
$y^{\mu}=y^{\mu}\left(
t,x^{i}\right)  $ and its conjugate momentum
$p_{\mu}=p_{\mu}\left(
t,x^{i}\right)  ,$ whose variation leads to the
kinematical equation (\ref{eq
din}).

Though from a mathematical point of view,
to fix the reference frame is, in
view of the reparameterization which restores the
canonical constraints, a
well defined procedure, it requires a physical interpretation;
indeed the open question is: which are the physical consequences
of fixing the slicing?\\ 
The complete answer to this question will be clear
at the end of this section,
but we can say already now that fixing the reference
frame we modify the system: the
dynamical equations and the constraints, which we can
obtain by the variation
of the action (\ref{azione totale}), describe no more
the dynamics of the
initial system formed by the gravitational and a
scalar field, but the coupled
system of these two fields with a non relativistic
dust fluid which interacts
with the field $A_{i}$ we have studied in the previous
section.
We remark that in a purely classical system it is not
necessary to introduce this additional term to the
gravitaty-matter action and therefore
we expect that the dust has not effect
on the dynamics of such systems have no undergone a classical
limit (but the non relativistic dust becomes important in the
description of those systems which evolve from a quantum state). 

Now to quantize the new constraints (\ref{new ham con}),
(\ref{new ham con2})
we use the canonical procedure, by implementing the canonical
variables to quantum operators,
i.e. $Q\rightarrow\widehat{Q},$ $P\rightarrow
\widehat{P};$ here $Q$ and $P$ denote, respectively,
a generic field and its conjugate momentum, which satisfy
the usual relations of commutation, i.e.

\begin{equation}
\left[  \widehat{Q}\left(  x\right)  ,\widehat{P}\left(  y\right)  \right]
=i\hbar\delta\left(  x-y\right)  ,
\end{equation}

in the functional representation we have

\begin{equation}
\widehat{Q}\left(  x\right)  \longrightarrow Q\left(  x\right)  ,\qquad
\widehat{P}\left(  y\right)  \longrightarrow-i\hbar\dfrac{\delta}{\delta
Q\left(  x\right)  },
\end{equation}

where the operators are applied to a wave functional
$\psi=\psi\left( Q\right)  .$

We assume that the state of the gravitational and
matter system be described
by a wave functional
$\Psi=\Psi\left(  y^{\mu},\phi,h_{ij},N,N^{i}\right)  $.
Then the new quantum dynamics of the whole system is
now described by the
functional differential system:%

\begin{align}
\dfrac{\delta\Psi}{\delta N}  &  =0,\qquad\qquad\qquad\qquad\qquad
\dfrac{\delta\Psi}{\delta N^{i}}=0,\\
i\hbar n^{\mu}\dfrac{\delta\Psi}{\delta y^{\mu}}=  &  \left(  \widehat{H}%
^{g}+\widehat{H}^{\phi}\right)  \Psi,\qquad i\hbar\partial_{i}y^{\mu}%
\dfrac{\delta\Psi}{\delta y^{\mu}}=\left(  \widehat{H}_{i}^{g}+\widehat{H}%
_{i}^{\phi}\right)  \Psi, \label{eq quantistiche}%
\end{align}

being
$\widehat{H}^{g}+\widehat{H}^{\phi}$and $\widehat{H}_{i}^{g}+\widehat
{H}_{i}^{\phi}$ the Hamiltonian operators after the quantum implementation of
the canonical variables. The first line of the above equations tell us that the wave functional
does not depend on the lapse function $N$ and the shift
vector $N^{i},$ so,
since now, we limit our attention on the other two equations,
considering that
the wave functional $\Psi$ depends only on the
3-metric $h_{ij}\left(
x^{k}\right)  $, the scalar
field $\phi\left(  x^{k}\right)  $ and the new
field $y^{\mu}\left(  x^{k}\right)  ,$ which
plays the role of the time
variable, by specifying the hypersurface on which the
wave functional is taken 
(we stress how its spatial gradients behaves like
potential terms).

Moreover, the second of
equation (\ref{eq quantistiche}) ensures the invariance
of the wave functional
under the spatial diffeomorphism and then, denoting by
the notation $\left\{
h_{ij}\right\}  $ a whole class of 3-geometries (i.e.
connected via 3-coordinates reparameterization),
the wave functional should be taken on such more appropriate
variable instead of a special realization of the 3-metric.\\
In the first of equations (\ref{eq quantistiche})
the vector field $n^{\mu }\left(  y^{\rho}\right)  $
is an arbitrary one without any peculiar
geometrical meaning; but when taking into account
the first of kinematical
equation (\ref{eq din}), $n^{\mu}$ becomes a real unit
normal vector field, since,
once fixed $N$ and $N^{i},$ $y^{\mu}\left(  t,x^{i}\right)  $
pays the price
for its geometrical interpretation.
These considerations lead us to claim that
the first of equation (\ref{eq din}) should be
included in the dynamics even
on the quantum level. The physical justification
for this statement
relies on the fact that no information about the
dynamic of the kinematical
dust comes from such an equation has discussed
in the previous section; in
fact there we have shown how the whole ``hydrodynamics'' of the dust be
entirely contained in the momentum equation.
In agreement to what we said in the introduction to this
work, the surviving of this classical equation on a quantum
level, reflects the classical nature of the ``device'' operating 
the (3 + 1)-splitting.\\ 
To take into account this
equation is equivalent to reduce $y^{\mu}$ to
a simple $\infty-$dimensional
parameter for the system dynamics.

In agreement with this point of view, we can smear the quantum dynamics on a
whole 1-parameter family of spatial hypersurfaces $\Sigma_{t}^{3}$ filling the
space-time; as soon as we introduce the notation
\begin{equation}
\partial_{t}=\underset{\Sigma_{t}^{3}}{\int}d^{3}x\partial_{t}y^{\mu}%
\dfrac{\delta}{\delta y^{\mu}},
\end{equation}

then equations (\ref{eq quantistiche}) acquire the Schr\"{o}dinger form
\begin{equation}
i\hbar\partial_{t}\Psi=\widehat{\mathcal{H}}\Psi, \label{raffaella3}%
\end{equation}
where

\begin{equation}
\widehat{\mathcal{H}}=\underset{\Sigma_{t}^{3}}{\int}d^{3}x\left[  N\left(
\widehat{H}^{g}+\widehat{H}^{\phi}\right)  +N^{i}\left(  \widehat{H}_{i}%
^{g}+\widehat{H}_{i}^{\phi}\right)  \right]  . \label{operatore hamiltoniano}%
\end{equation}

In this new framework the wave functional can be taken directly on the label
time $\left(  \text{i.e. }\Psi=\Psi\left(  t,\phi,h_{ij}\right)  \right)$ (where we have removed the curl bracket from $h_{ij}$ because, when regarded in such a manner, the wave functional is no longer invariant under 3-diffeomorphism), 
since the latter becomes a physical clock via the correspondence, we show
below, between the eigenvalue problem of the equation (\ref{raffaella3}) and
the energy-momentum of the dust discussed in the previous section.

In order to construct the Hilbert space associated to the Schr\"{o}dinger-like
equation we must prove the Hermitianity of the Hamiltonian operator; since
the Hermitianity of the $\phi$ term was proven in \cite{Kuc1981}, as well as
the Hermitianity of the operator $\widehat{H}^{g}$ in \cite{Mon2002} under the
following choice for the normal ordering
\begin{equation}
G_{ijkl}p^{ij}p^{kl}\rightarrow-\hbar^{2}\dfrac{\delta}{\delta h_{ij}}\left(
G_{ijkl}\dfrac{\delta\left(  ...\right)  }{\delta h_{kl}}\right)  ,
\label{ordinamento}%
\end{equation}

then it remains to be shown the Hermitianity of the operator $\widehat
{h}=\underset{\Sigma_{t}^{3}}{\int}d^{3}xN^{i}\widehat{H}_{i}^{g}$. In Dirac
notation we have to show that:
\begin{equation}
\left\langle \Psi_{1}\left|  \widehat{h}\right|  \Psi_{2}\right\rangle
=\left\langle \Psi_{2}\left|  \widehat{h}\right|  \Psi_{1}\right\rangle
^{\ast}.
\end{equation}

To this aim we write down the explicit expression of the above bracket:
\begin{equation}
\left\langle \Psi_{1}\left|  \widehat{h}\right|  \Psi_{2}\right\rangle
=2i\hbar%
%TCIMACRO{\dint \limits_{\mathcal{F}_{t}}}%
%BeginExpansion
{\displaystyle\int\limits_{\mathcal{F}_{t}}}
%EndExpansion
Dh\underset{\Sigma_{t}^{3}}{\int}d^{3}x\Psi_{1}^{\ast}N^{i}h_{ik}\nabla
_{j}\dfrac{\delta}{\delta h_{kj}}\Psi_{2}, \label{braket}%
\end{equation}

where $Dh$ is the Lebesgue measure in the 3-geometries functional space.

Now integrating by parts, considering that the hypersurfaces $\Sigma_{t}^{3}$
are compact and using, in view of the functional Gauss theorem, the following
relation:%
\begin{equation}%
%TCIMACRO{\dint \limits_{\mathcal{F}_{t}}}%
%BeginExpansion
{\displaystyle\int\limits_{\mathcal{F}_{t}}}
%EndExpansion
Dh\underset{\Sigma_{t}^{3}}{%
%TCIMACRO{\dint }%
%BeginExpansion
{\displaystyle\int}
%EndExpansion
}d^{3}x\dfrac{\delta}{\delta h_{kj}}\left(  ......\right)=0 \label{gauss},
\end{equation}

we can rewrite the expression (\ref{braket}) in the following form:%
\begin{equation}
\left\langle \Psi_{1}\left|  \widehat{h}\right|  \Psi_{2}\right\rangle
=2i\hbar%
%TCIMACRO{\dint \limits_{\mathcal{F}_{t}}}%
%BeginExpansion
{\displaystyle\int\limits_{\mathcal{F}_{t}}}
%EndExpansion
Dh\underset{\Sigma_{t}^{3}}{%
%TCIMACRO{\dint }%
%BeginExpansion
{\displaystyle\int}
%EndExpansion
}d^{3}x\dfrac{\delta}{\delta h_{kj}}\left(  \Psi_{1}^{\ast}\left(  \nabla
_{j}N^{i}\right)  h_{ik}\right)  \Psi_{2}. \label{integr}%
\end{equation}

It is possible to show that two of the terms, which come from the right side
of (\ref{integr}) when the functional derivative operates on the quantities in
the parenthesis, are zero. In fact, acting with the functional derivative on
the 3-metric, we obtain:%
\begin{equation}
2i\hbar%
%TCIMACRO{\dint \limits_{\mathcal{F}_{t}}}%
%BeginExpansion
{\displaystyle\int\limits_{\mathcal{F}_{t}}}
%EndExpansion
Dh\underset{\Sigma_{t}^{3}}{%
%TCIMACRO{\dint }%
%BeginExpansion
{\displaystyle\int}
%EndExpansion
}d^{3}x\Psi_{1}^{\ast}\left(  \nabla_{j}N^{i}\right)  \dfrac{\delta h_{ik}%
}{\delta h_{kj}}\Psi_{2}=-2i\hbar%
%TCIMACRO{\dint \limits_{\mathcal{F}_{t}}}%
%BeginExpansion
{\displaystyle\int\limits_{\mathcal{F}_{t}}}
%EndExpansion
Dh\underset{\Sigma_{t}^{3}}{%
%TCIMACRO{\dint }%
%BeginExpansion
{\displaystyle\int}
%EndExpansion
}d^{3}x\nabla_{j}\left(  \Psi_{1}^{\ast}\Psi_{2}\right)  N^{j}, \label{zero}%
\end{equation}

where we have integrated by parts and used the compactness of the
hypersurfaces $\Sigma_{t}^{3}.$ But the right hand side of (\ref{zero}) is
zero, because $\Psi$ is a functional, so it does not depend on $x.$

When the functional derivative in expression (\ref{integr}) acts on the
covariant derivative of the shift vector, we obtain:%

\begin{align}
  2i\hbar%
%TCIMACRO{\dint \limits_{\mathcal{F}_{t}}}%
%BeginExpansion
{\displaystyle\int\limits_{\mathcal{F}_{t}}}
%EndExpansion
Dh\underset{\Sigma_{t}^{3}}{%
%TCIMACRO{\dint }%
%BeginExpansion
{\displaystyle\int}
%EndExpansion
}&d^{3}xh_{ik}\Psi_{1}^{\ast}\Psi_{2}\dfrac{\delta}{\delta h_{kj}}\left(
\nabla_{j}N^{i}\right)  =\nonumber\\
&  =2i\hbar%
%TCIMACRO{\dint \limits_{\mathcal{F}_{t}}}%
%BeginExpansion
{\displaystyle\int\limits_{\mathcal{F}_{t}}}
%EndExpansion
Dh\underset{\Sigma_{t}^{3}}{%
%TCIMACRO{\dint }%
%BeginExpansion
{\displaystyle\int}
%EndExpansion
}d^{3}xh_{ik}\Psi_{1}^{\ast}\Psi_{2}\dfrac{\delta}{\delta h_{kj}}\left(
\Gamma_{jm}^{i}N^{m}\right)  , \label{zero2}%
\end{align}

since in the right side term, the derivative operator is applied to a function
of $x$ and not to a functional, thus, like in the case of the variation with
respect a dynamical variable, the ordinary derivative operator and the
functional one commute, so it is simple to show that $\dfrac{\delta}{\delta
h_{kj}}\left(  \Gamma_{jm}^{i}N^{m}\right)  =0,$ thus the term (\ref{zero2})
is identically zero.

Finally the expression (\ref{integr}) can be rewrite:%
\begin{align}
\left\langle \Psi_{1}\left|  \widehat{h}\right|  \Psi_{2}\right\rangle  &
=2i\hbar%
%TCIMACRO{\dint \limits_{\mathcal{F}_{t}}}%
%BeginExpansion
{\displaystyle\int\limits_{\mathcal{F}_{t}}}
%EndExpansion
Dh\underset{\Sigma_{t}^{3}}{%
%TCIMACRO{\dint }%
%BeginExpansion
{\displaystyle\int}
%EndExpansion
}d^{3}x\dfrac{\delta\Psi_{1}^{\ast}}{\delta h_{kj}}\left(  \nabla_{j}%
N^{i}\right)  h_{ik}\Psi_{2}=\nonumber\\
&  =-2i\hbar%
%TCIMACRO{\dint \limits_{\mathcal{F}_{t}}}%
%BeginExpansion
{\displaystyle\int\limits_{\mathcal{F}_{t}}}
%EndExpansion
Dh\underset{\Sigma_{t}^{3}}{%
%TCIMACRO{\dint }%
%BeginExpansion
{\displaystyle\int}
%EndExpansion
}d^{3}x\Psi_{2}N^{i}h_{ik}\nabla_{j}\dfrac{\delta\Psi_{1}^{\ast}}{\delta
h_{kj}}=\left\langle \Psi_{2}\left|  \widehat{h}\right|  \Psi_{1}\right\rangle
^{\ast}.
\end{align}

The above equality assures $\widehat{\mathcal{H}}$ is an Hermitian operator.

Defining the following inner product:
\begin{equation}
\left\langle \Psi_{1}\mid\Psi_{2}\right\rangle =\underset{y_{t}}{\int}%
DhD\phi\Psi_{1}^{\ast}\Psi_{2}, \label{bracket}%
\end{equation}

where $DhD\phi$ is the Lebesgue measure for the functional space of all the
dynamical variables and $y_{t}$ is the corresponding functional domain, we can
turn the space of solutions of the Schr\"{o}dinger-like equation into an
Hilbert space. We interpret the above bracket as the probability that a state
$\left|  \Psi_{1}\right\rangle $ falls into another state $\left|  \Psi
_{2}\right\rangle $ and, defining the density of probability $\rho=\Psi^{\ast
}\Psi,$ we can also construct the amplitude for the system lying in a field
configuration. By the Hermitian character of the operator $\widehat
{\mathcal{H}},$ it is possible to show that the probability is constant in
time, in fact:
\begin{equation}
\partial_{t}\left\langle \Psi_{1}\right|  \left.  \Psi_{2}\right\rangle
=\underset{\Sigma_{t}^{3}}{%
%TCIMACRO{\dint }%
%BeginExpansion
{\displaystyle\int}
%EndExpansion
}d^{3}x\partial_{t}y^{\mu}\dfrac{\delta}{\delta y^{\mu}}\left\langle \Psi
_{1}\right|  \left.  \Psi_{2}\right\rangle =\dfrac{i}{\hbar}\left(
\left\langle \widehat{\mathcal{H}}\Psi_{1}\mid\Psi_{2}\right\rangle
-\left\langle \Psi_{1}\mid\widehat{\mathcal{H}}\Psi_{2}\right\rangle \right)
=0,
\end{equation}

and the general character of the deformation vector allows us to write the
fundamental conservation law
\begin{equation}
\dfrac{\delta\left\langle \Psi_{1}\right|  \left.  \Psi_{2}\right\rangle
}{\delta y^{\mu}}=0,
\end{equation}

which assures the probability does not depend on the choice of the hypersurface.

The density of probability $\rho$ satisfies a continuity equation, which can
be obtained multiplying the Schr\"{o}dinger-like equation times the complex
conjugate wave function $\Psi^{\ast}$ and the complex conjugate equation times the wave function $\Psi,$ i.e.
\begin{equation}
i\hbar\Psi^{\ast}\partial_{t}\Psi=\Psi^{\ast}\widehat{\mathcal{H}}\Psi,
\qquad-i\hbar\Psi\partial_{t}\Psi^{\ast}=\Psi\widehat{\mathcal{H}}^{\ast}%
\Psi^{\ast}, \label{equazioni sopra}%
\end{equation}

subtracting the second of equation (\ref{equazioni sopra}) from the first one,
we obtain:%
\begin{align}
i\hbar\partial_{t}\left(  \Psi\Psi^{\ast}\right)   &  =\underset{\Sigma
_{t}^{3}}{%
%TCIMACRO{\dint }%
%BeginExpansion
{\displaystyle\int}
%EndExpansion
}d^{3}x\left\{  -\hbar^{2}\left(  \Psi^{\ast}N\dfrac{\delta}{\delta h_{ij}%
}G_{ijkl}\dfrac{\delta}{\delta h_{kl}}\Psi-\Psi N\dfrac{\delta}{\delta h_{ij}%
}G_{ijkl}\dfrac{\delta}{\delta h_{kl}}\Psi^{\ast}\right)  \right.
+\nonumber\\
&  -\hbar^{2}\left(  \Psi^{\ast}\dfrac{N}{2\sqrt{h}}\dfrac{\delta}{\delta\phi
}\dfrac{\delta}{\delta\phi}\Psi-\Psi\dfrac{N}{2\sqrt{h}}\dfrac{\delta}%
{\delta\phi}\dfrac{\delta}{\delta\phi}\Psi^{\ast}\right)  +\nonumber\\
&  +2i\hbar\left(  \Psi^{\ast}N^{i}h_{ik}\nabla_{j}\dfrac{\delta}{\delta
h_{kj}}\Psi+\Psi N^{i}h_{ik}\nabla_{j}\dfrac{\delta}{\delta h_{kj}}\Psi^{\ast
}\right)  +\nonumber\\
&  -\left.  i\hbar\left(  \Psi^{\ast}N^{i}\partial_{i}\phi\dfrac{\delta
}{\delta\phi}\Psi+\Psi N^{i}\partial_{i}\phi\dfrac{\delta}{\delta\phi}%
\Psi^{\ast}\right)  \right\}  , \label{eq cont}%
\end{align}

defining now the tensor probability current $A_{ij}$, which is connected with
the 3-metric tensor field, in the following way:%
\begin{equation}
A_{ij}=-i\hbar\left(  \Psi^{\ast}NG_{ijkl}\dfrac{\delta}{\delta h_{kl}}%
\Psi-\Psi NG_{ijkl}\dfrac{\delta}{\delta h_{kl}}\Psi^{\ast}\right)
+2h_{ki}\left(  \nabla_{j}N^{k}\right)  \Psi^{\ast}\Psi,
\end{equation}

and the scalar probability current $A,$ connected, instead, to the presence of
the scalar field $\phi,$ as:%

\begin{equation}
A=-i\hbar\dfrac{N}{2\sqrt{h}}\left(  \Psi^{\ast}\dfrac{\delta}{\delta\phi}%
\Psi-\Psi\dfrac{\delta}{\delta\phi}\Psi^{\ast}\right)  -i\hbar\left(
\phi\partial_{i}N^{i}\Psi^{\ast}\Psi\right)  ,
\end{equation}

the equation (\ref{eq cont}) takes the following form:
\begin{equation}
\partial_{t}\rho+\underset{\Sigma_{t}^{3}}{%
%TCIMACRO{\dint }%
%BeginExpansion
{\displaystyle\int}
%EndExpansion
}d^{3}x\left(  \dfrac{\delta A_{ij}}{\delta h_{ij}}+\dfrac{\delta A}%
{\delta\phi}\right)  =0,
\end{equation}

integrating on the functional space $y_{t}$, using the generalized Gauss theorem (\ref{gauss}), the continuity equation assures
that the probability is constant in time as above.

Let us now reconsider the Schr\"{o}dinger dynamics in terms of a time
independent eigenvalues problem. To this end we expand the wave functional as
follows:
\begin{align}
\Psi\left(  t,\phi,h_{ij}\right)=\underset{y_{t}^{\ast}}{%
%TCIMACRO{\dint }%
%BeginExpansion
{\displaystyle\int}
%EndExpansion
}D\Omega&DK\Theta\left(  \Omega,K_{i}\right)  \chi_{\Omega,K_{i}}\left(
\phi,h_{ij}\right)  \cdot\nonumber\\
&  \cdot\exp\left\{  -\dfrac{i}{\hbar}\int\limits_{t_{0}}^{t}dt^{\prime}%
\underset{\Sigma_{t}^{3}}{\int}d^{3}x\left(  N\Omega+N^{i}K_{i}\right)
\right\},  \label{expan}%
\end{align}

being $t_{0}$ an assigned initial ``instant''. Where $D\Omega DK$ denotes the
Lebesgue measure in the functional space $y_{t}^{\ast}$ of the conjugate
function $\Omega\left(  x^{i}\right)  $ and $K_{i}\left(  x^{i}\right)  ,$
$\Theta=\Theta\left(  \Omega,K_{i}\right)  $ a functional valued in this
domain, whose form is determined by the initial conditions $\Psi_{0}%
=\Psi\left(  t_{0},\phi,h_{ij}\right)  $. When we substitute the expansion
(\ref{expan}) of the wave functional into the equation (\ref{raffaella3}), it
is satisfied only if take place the following $\infty^{3}-$dimensional
eigenvalues problem:

\begin{equation}
\left(  \widehat{H}^{g}+\widehat{H}^{\phi}\right)  \chi_{_{\Omega,K_{i}}%
}=\Omega\left(  x^{j}\right)  \chi_{_{\Omega,K_{i}}},\qquad \left(  \widehat
{H}_{i}^{g}+\widehat{H}_{i}^{\phi}\right)  \chi_{_{\Omega,K_{i}}}=K_{i}\left(
x^{j}\right)  \chi_{_{\Omega,K_{i}}}. \label{eigenvalue}%
\end{equation}

Now to characterize the physical meaning of the above eigenvalues, we
construct the semi-classical limit of the Schr\"{o}dinger-like equation, by
splitting the wave functional into its modulus and phase, as follows:
\begin{equation}
\Psi=\sqrt{\rho}e^{\tfrac{i}{\hbar}\sigma} \label{psi semiclassical}%
\end{equation}

Then in the limit $\hbar\rightarrow0$ we obtain for $\sigma$ an
Hamilton-Jacobi equation of the form:
\begin{align}
-\partial_{t}\sigma &  =\underset{\Sigma_{t}^{3}}{%
%TCIMACRO{\dint }%
%BeginExpansion
{\displaystyle\int}
%EndExpansion
}d^{3}xN\left(  G_{ijkl}\dfrac{\delta\sigma}{\delta h_{ij}}\dfrac{\delta
\sigma}{\delta h_{ij}}-\sqrt{h}^{\left(  3\right)  }R\right.  +\nonumber\\
&  +\left.  \dfrac{1}{2\sqrt{h}}\dfrac{\delta\sigma}{\delta\phi}\dfrac
{\delta\sigma}{\delta\phi}+\dfrac{\sqrt{h}}{2}h^{ij}\partial_{i}\phi
\partial_{j}\phi+\sqrt{h}V\left(  \phi\right)  \right)  +\nonumber\\
&  -\underset{\Sigma_{t}^{3}}{%
%TCIMACRO{\dint }%
%BeginExpansion
{\displaystyle\int}
%EndExpansion
}d^{3}xN^{i}\left(  2h_{ik}\nabla_{j}\dfrac{\delta\sigma}{\delta h_{kj}%
}-\partial_{i}\phi\dfrac{\delta\sigma}{\delta\phi}\right)  \label{ham-jac}%
\end{align}

The non vanishing of the $\sigma$ time derivative reflects the evolutive
character appearing in the constructed theory and makes account for the
presence, on the classical limit, of the dust matter discussed in the previous
section. To clarify this feature, we set
\begin{equation}
\sigma\left(  t,\phi,h_{ij}\right)  =\tau\left(  \phi,h_{ij}\right)
+\int\limits_{t_{0}}^{t}dt^{\prime}%
\underset{\Sigma_{t}^{3}}{%
%TCIMACRO{\dint }%
%BeginExpansion
{\displaystyle\int}
%EndExpansion
}d^{3}x\left(  N\Omega+N^{i}K_{i}\right)  .
\end{equation}

When we substitute this expression in the Hamilton-Jacobi equation, and
identify $p^{ij}=\dfrac{\delta\tau}{\delta h_{ij}},$ $p_{\phi}%
=\dfrac{\delta\tau}{\delta\phi},$ then the equation (\ref{ham-jac}) becomes
equivalent to the $\infty-$dimensional ones:

\begin{equation}
\left(  H^{g}+H^{\phi}\right)=\Omega\left(  x^{j}\right)\qquad \left(H_{i}^{g}+H_{i}^{\phi}\right)=K_{i}\left(  x^{j}\right)
\end{equation}

We stress how these equations coincides with those ones obtainable by the
eigenvalues problem (\ref{eigenvalue}), as soon as we choose the classical
limit of $\chi\sim e^{\tfrac{i}{\hbar}\tau}$ thus, at the end of this
analysis, recalling expressions (\ref{raffaella}) and (\ref{new ham con2}), we can identify the super-Hamiltonian eigenvalue $\Omega$ with $\omega$
and the super-momentum eigenvalues $K_{i}$ with $k_{i}.$ On the other hand by
equations (\ref{raffaella}) and (\ref{ham mom2}), the above identification
implies that: $\Omega=-\sqrt{h}\varepsilon$ and $K_{i}=-\gamma A_{i}\omega.$

The relation we obtained show how super-Hamiltonian and super-momentum
eigenvalues are directly connected with the dust fields introduced in section
\ref{par2}. Even starting from a quantum point of view we recognize the existence of a
dust fluid playing the role of a physical clock for the gravity-matter dynamics.

\bigskip

\section{The Ehrenfest theorem}

\label{4}

\bigskip

In this section we face the problem to derive an Ehrenfest approach for the
expectation values dynamics, to this end we start by the natural relation:
\begin{equation}
\dfrac{d}{dt}\left\langle \widehat{T}\right\rangle =\dfrac{1}{i\hbar
}\left\langle \left[  \widehat{T},\widehat{\mathcal{H}}\right]  \right\rangle
\end{equation}

where $\widehat{T}$ denotes a generic operator, acting on physical states. Now
we implement this formula for the case of the 3-metric $\widehat{h}_{ij}$ and
its conjugate momentum $\widehat{p}^{ij}$ as well as for the scalar field
$\phi$ and its conjugate momentum $p_{\phi}$; in view of the canonical equal
time commutation relations:%

\begin{equation}
\left[  \phi,F\left(  \phi\right)  \right]  =\left[  p_{\phi},G\left(
p_{\phi}\right)  \right]  =0
\end{equation}%

\begin{equation}
\left[  \phi\left(  x\right)  ,p_{\phi}\left(  y\right)  \right]
=i\hbar\delta\left(  x-y\right)
\end{equation}%
\begin{equation}
\left[  \widehat{h}_{ij}\left(  x\right)  ,F\left(  \widehat{h}_{kl}\left(
y\right)  \right)  \right]  =\left[  \widehat{p}^{ij}\left(  x\right)
,G\left(  \widehat{p}^{kl}\left(  y\right)  \right)  \right]  =0
\end{equation}%

\begin{equation}
\left[  \widehat{h}_{ij}\left(  x\right)  ,\widehat{p}^{kl}\left(  y\right)
\right]  =i\hbar\delta_{ij}^{kl}\delta\left(  x-y\right)
\end{equation}

where $F$ and $G$ are two generic function; we obtain the following equations
for the expectation value of the 3-metric field:
\begin{align}
\dfrac{d}{dt}\left\langle \widehat{h}_{ab}\left(  x\right)  \right\rangle  &
=\dfrac{1}{i\hbar}%
%TCIMACRO{\dint }%
%BeginExpansion
{\displaystyle\int}
%EndExpansion
d^{3}y\left\{  N\left\langle \left[  \widehat{h}_{ab}\left(  x\right)
,\widehat{p}^{ij}\left(  y\right)  \right]  \widehat{G}_{ijkl}\left(
y\right)  \widehat{p}^{kl}\left(  y\right)  \right\rangle \right.
+\nonumber\\
&  +N\left\langle \widehat{p}^{ij}\left(  y\right)  \widehat{G}\left(
y\right)  _{ijkl}\left[  \widehat{h}_{ab}\left(  x\right)  ,\widehat{p}%
^{kl}\left(  y\right)  \right]  \right\rangle +\nonumber\\
&  +\left.  2\nabla_{j}N_{i}\left\langle \left[  \widehat{h}_{ab}\left(
x\right)  ,\widehat{p}^{ij}\left(  y\right)  \right]  \right\rangle \right\}
.
\end{align}

Now expliciting the commutators, the above expression rewrites as:
\begin{align}
\dfrac{d}{dt}\left\langle \widehat{h}_{ab}\left(  x\right)  \right\rangle  &
=N\left\langle \left(  \widehat{G}_{abkl}\left(  x\right)  \widehat{p}%
^{kl}\left(  x\right)  +\widehat{p}^{ij}\left(  x\right)  \widehat{G}\left(
x\right)  _{ijab}\right)  \right\rangle +\nonumber\\
&  +2\left\langle \nabla_{(a}N_{b)}\right\rangle .
\end{align}

Using the commutation relation, after few algebra, we obtain:
\begin{align}
\dfrac{d}{dt}\left\langle \widehat{h}_{ab}\left(  x\right)  \right\rangle  &
=2N\left\langle \widehat{G}_{abkl}\left(  x\right)  \widehat{p}^{kl}\left(
x\right)  \right\rangle +\nonumber\\
&  +2\left\langle \nabla_{(a}N_{b)}\right\rangle -i\hbar\dfrac{3}%
{4}N\left\langle \left(  \widehat{h}\left(  x\right)  \right)  ^{-1/2}%
\widehat{h}_{ab}\left(  x\right)  \right\rangle
\end{align}

For what regards the conjugate momentum operator, we have:
\begin{align}
&  \dfrac{d}{dt}\left\langle \widehat{p}^{ab}\left(  x\right)  \right\rangle
=\dfrac{1}{i\hbar}%
%TCIMACRO{\dint }%
%BeginExpansion
{\displaystyle\int}
%EndExpansion
d^{3}y\left\{  N\left\langle \widehat{p}^{ij}\left(  y\right)  \left[
\widehat{p}^{ab}\left(  x\right)  ,\widehat{G}_{ijkl}\left(  y\right)
\right]  \widehat{p}^{kl}\left(  y\right)  \right\rangle \right.  +\nonumber\\
&  -N\left\langle \left[  \widehat{p}^{ab}\left(  x\right)  ,\widehat{h}%
^{1/2}\left(  y\right)  ^{(3)}\widehat{R}\left(  y\right)  \right]
\right\rangle +2\left\langle \widehat{p}^{ij}\left(  y\right)  \left[
\widehat{p}^{ab}\left(  x\right)  ,\nabla_{j}\right]  N_{i}\right\rangle
+\nonumber\\
&  +\dfrac{N}{2}\left\langle \widehat{p}_{\phi}^{2}\left(  y\right)  \left[
\widehat{p}^{ab}\left(  x\right)  ,\widehat{h}^{-1/2}\left(  y\right)
\right]  \right\rangle +\nonumber\\
&  +N\left\langle \left(  \dfrac{1}{2}h^{ij}\left(  y\right)  \partial
_{i}\widehat{\phi}\left(  y\right)  \partial_{j}\widehat{\phi}\left(
y\right)  +\widehat{V}\left(  \phi\right)  \right)  \left[  \widehat{p}%
^{ab}\left(  x\right)  ,\widehat{h}^{1/2}\left(  y\right)  \right]
\right\rangle
\end{align}

expliciting the commutators, with a certain algebra, we obtain:%

\begin{align}
\dfrac{d}{dt}\left\langle p^{ab}\right\rangle  &  =\dfrac{1}{2}\left\langle
N\widehat{h}^{-1/2}\widehat{h}^{ab}\left(  \widehat{p}^{ij}\widehat{p}%
_{ij}-\dfrac{1}{2}\widehat{p}^{2}\right)  \right\rangle -\left\langle
2N\widehat{h}^{-1/2}\left(  \widehat{p}^{ai}\widehat{p}_{i}^{b}-\dfrac{1}%
{2}\widehat{p}\widehat{p}^{ab}\right)  \right\rangle +\nonumber\\
&  -\left\langle N\widehat{h}^{1/2}\left(  ^{\left(  3\right)  }\widehat
{R}^{ab}-\dfrac{1}{2}^{\left(  3\right)  }\widehat{R}\widehat{h}^{ab}\right)
\right\rangle +\left\langle \widehat{h}^{1/2}\left(  \nabla^{a}\nabla
^{b}N-\widehat{h}^{ab}\nabla^{i}\nabla_{i}N\right)  \right\rangle +\nonumber\\
&  -\left\langle 2\nabla_{i}\left(  \widehat{p}^{i(a}N^{b)}\right)
+\nabla_{i}\left(  N^{i}\widehat{p}^{ab}\right)  \right\rangle +\nonumber\\
&  +\left\langle \dfrac{1}{4}\widehat{h}^{-1/2}\widehat{h}^{ab}\widehat
{p}_{\phi}^{2}-\dfrac{1}{2}\widehat{h}^{1/2}\widehat{h}^{ab}\left(  \dfrac
{1}{2}h^{ij}\partial_{i}\widehat{\phi}\partial_{j}\widehat{\phi}+\widehat
{V}\left(  \phi\right)  \right)  \right\rangle +\nonumber\\
&  -i\hbar\dfrac{3}{4}\left\langle N\widehat{h}^{-1/2}\left(  \widehat{p}%
^{ab}-\dfrac{1}{2}\widehat{h}^{ab}\widehat{p}\right)  \right\rangle .
\end{align}

\bigskip Now to obtain the dynamical equations for the mean values of the
scalar field $\phi$ and its conjugate momentum $p_{\phi}$, we write like
above:
\begin{align}
\dfrac{d}{dt}\left\langle \widehat{\phi}\left(  x\right)  \right\rangle  &
=\dfrac{1}{i\hbar}%
%TCIMACRO{\dint }%
%BeginExpansion
{\displaystyle\int}
%EndExpansion
d^{3}y\dfrac{1}{2}\left\{  \left\langle N\widehat{h}^{-1/2}\left(  y\right)
\left[  \widehat{\phi}\left(  x\right)  ,\widehat{p}_{\phi}^{2}\left(
y\right)  \right]  \right\rangle \right.  +\nonumber\\
&  +\left.  \left\langle N^{i}\left[  \widehat{\phi}\left(  x\right)
,\widehat{p}_{\phi}\left(  y\right)  \right]  \partial_{i}\widehat{\phi
}\left(  y\right)  \right\rangle \right\}  ,
\end{align}

using the commutation relations we simply obtain:
\begin{equation}
\dfrac{d}{dt}\left\langle \widehat{\phi}\left(  x\right)\right\rangle =\left\langle
N\widehat{h}^{-1/2}\left(  x\right)\widehat{p}_{\phi}\left(  x\right)\right\rangle +\left\langle N^{i}%
\partial_{i}\widehat{\phi}\left(  x\right)\right\rangle .
\end{equation}

For what regards the conjugate momentum of the scalar field we have%

\begin{align}
&  \dfrac{d}{dt}\left\langle \widehat{p}_{\phi}\left(  x\right)  \right\rangle
=\dfrac{1}{i\hbar}%
%TCIMACRO{\dint }%
%BeginExpansion
{\displaystyle\int}
%EndExpansion
d^{3}y\left\{  \left\langle \partial_{i}\left(  -\dfrac{1}{2}N\widehat
{h}^{1/2}\widehat{h}^{ij}\left(  y\right)  \partial_{j}\widehat{\phi}\left(
y\right)  \right)  \left[  \widehat{p}_{\phi}\left(  x\right)  ,\widehat{\phi
}\left(  y\right)  \right]  \right\rangle \right.  +\nonumber\\
&  +\left.  \left\langle N\widehat{h}^{1/2}\left(  y\right)  \left[
\widehat{p}_{\phi}\left(  x\right)  ,\widehat{V}\left(  \phi\right)  \right]
\right\rangle -\left\langle \partial_{i}\left(  N^{i}\widehat{p}_{\phi}\left(
y\right)  \right)  \left[  \widehat{p}_{\phi}\left(  x\right)  ,\widehat{\phi
}\left(  y\right)  \right]  \right\rangle \right\}  ,
\end{align}

by which the dynamical equation below follows
\begin{align}
\dfrac{d}{dt}\left\langle p_{\phi}\left(  x\right)\right\rangle  &  =\left\langle N\widehat
{h}^{1/2}\left(  x\right)\widehat{h}^{ij}\left(  x\right)\partial_{i}\partial_{j}\widehat{\phi}\left(  x\right)\right\rangle
+\left\langle \partial_{j}\left(  N\widehat{h}^{1/2}\left(  x\right)\widehat{h}^{ij}\left(  x\right)\right)
\partial_{i}\widehat{\phi}\left(  x\right)\right\rangle +\nonumber\\
&  -\left\langle N\widehat{h}^{1/2}\left(  x\right)\dfrac{\partial V\left(  \phi\right)
}{\partial\phi}\right\rangle +\left\langle \partial_{i}\left(  N^{i}%
\widehat{p}_{\phi}\left(  x\right)\right)  \right\rangle .
\end{align}

We observe that the equations we obtain for the expectation values of the
3-metric field and of its conjugate momentum differ from those classical ones
(\ref{eq per h}) and (\ref{eq per p}), by a term which is due to the
particular normal ordering, we must choose in the theory. However in the
semiclassical limit $\hbar\rightarrow0$, when the wave function $\Psi$ is
taken in the form (\ref{psi semiclassical}), such terms vanish because do not
contain functional derivatives acting on the wave functional. So in this limit
we get exactly the classical equations, if the function $\rho$ approaches to
$\delta$ function and if the canonical variables are obtained from the
Hamilton-Jacobi principal function. By other words the additional terms in
$\hbar$ vanish in the semiclassical limit simply because, in such condition,
to speak of normal ordering makes no longer sense.

It is worth noting that the Ehrenfest procedure here discussed does not allow
to reproduce the expectation values for the Hamiltonian constraints, but this
is a natural consequence of fixing the lapse function $N$ and the shift vector
$N^{i}$ in the gravity-matter action. The impossibility to have the vanishing
Hamiltonian expectation values in this way leads just to the quantum
phenomenology for the dust energy-momentum.

\bigskip

\section{Concluding remarks}

\bigskip

We have presented a reformulation of the canonical
quantization of
geometrodynamics with respect
to a fixed reference frame;
the main result
obtained, including the kinematical action
in the global dynamics, is the
characterization of an appropriate internal
physical clock. In our theory the
role of clock is played by the reference fluid,
comoving with the 3-hypersurfaces
and its presence is necessary to distinguish between
space-like and time-like geometrical objects
before the canonical quantization procedure.

The fluid shows its presence through a comoving
(non-positive defined) density of energy and momentum, which we
have characterized either from a classical
either from a quantum point of
view: classically it comes from
having introduced the kinematical action, but
its real nature must be investigated in the
classical limit of the eigenvalues equations.

Some aspects in this theory need to be
further developed on to be completely
understood. We think of fundamental importance
be the study of the
phenomenological implications which this theory
can have in a cosmological
level; indeed the density of the dust
and the field $A_{i}$ could have direct
something to do with the
\emph{dark matter } component observed in the Universe.

To be applicable to a generic inhomogeneous gravitational
system, the theory here presented has to
be reduced, necessarily, to a formulation on a lattice;
recently some interesting proposal has appeared
to discretize a quantum constraint
\cite{DiBGamPul2002}, \cite{GamPul2002}and
they are of course relevant for the
discretization of the present theory. A more direct
approach can be obtained
applying the Regge calculus \cite{Reg1961}, \cite{Reg1997},
to the 3-geometries on the spatial hypersurfaces.

\end{document}